\documentstyle[12pt]{article}
\newcommand{\ice}[1]{#1}
\newcommand{\beq}{\begin{equation}}
\newcommand{\eeq}{\end{equation}}
\newcommand{\api}{\frac{\alpha_s}{\pi}}

\newcommand{\ba}{\begin{array}}
\newcommand{\ea}{\end{array}}

\newcommand{\as}{\alpha_s}

\newcommand{\g}{\gamma}

\newcommand{\msbar}{\overline{\mbox{MS}}}
\newcommand{\ordas}{{\cal O}(\alpha_s}
\newcommand{\dsp}{\displaystyle}
\newcommand{\EQN}{\label}
\newcommand{\ovl}{\overline}
\newcommand{\dif}{{\rm d}}
\newcommand{\ex}{{\rm e}}
\newcommand{\fos}[2]{\>\>\mathop{#2}^{{}\atop{#1}}{}}
\def\bbuildrel#1_#2^#3%
{\mathrel{\mathop{\kern 0pt#1}\limits_{#2}^{#3}}}

\begin{document}

\begin{titlepage}
\noindent
%
%
\hfill TTP94--10\\
\mbox{}
\hfill  July 1994   \\
%
%
\vspace{-0.5cm}
\begin{center}
  \begin{Large}
  \begin{bf}
Perturbative QCD Corrections to the $Z$ Boson
Width and the Higgs Decay Rate
  \end{bf}
  \end{Large}
\footnote{
Summary of contributed talks at

1) XXIXth Rencontres de Moriond, QCD and High Energy Interactions,

 M\'eribel, March 19 -- 26, 1994

2) Workshop: QCD at LEP, Aachen, April 11, 1994

3) Zeuthen Workshop on Elementary Particle Theory:
Physics at LEP200 and Beyond,

 Teupitz, April 10 -- 15, 1994.}

%
%
  \vspace{0.5cm}
  \begin{large}
K.G.Chetyrkin  \\
Institute for Nuclear Research\\
Russian Academy of Sciences\\
60th October Anniversary Prospect 7a\\
 Moscow, 117312, Russia \\[5mm]
 J.H.K\"uhn, A.Kwiatkowski    \\
    Institut f\"ur Theoretische Teilchenphysik \\
    Universit\"at Karlsruhe \\
    D-76128 Karlsruhe, Germany\\
  \end{large}
%
%
  \vspace{0.5cm}
  {\bf Abstract}
\end{center}
\begin{quotation}

\noindent
Radiative QCD corrections
 significantly influence the theoretical
predictions for the decay rates
of the $Z$  and the Higgs boson.
The status of the QCD calculations to the
hadronic $Z$ width is reviewed.
The role of mass corrections from bottom
quark final states is emphasized.
An estimate of the theoretical uncertainties
is given.
New results for quartic mass terms of order
${\cal O}(\alpha_s^2)$ are presented. The impact
of secondary radiation of bottom quarks on the
determination of $\Gamma(Z\rightarrow b\bar{b})$
is discussed.
Second order QCD corrections to the partial
decay rate $\Gamma(H\rightarrow b\bar{b})$
are also presented in this talk. A recent result
for the flavour singlet contribution to this
quantity is presented. It includes quark mass
effects and completes the otherwise massless
calculations of order $\ordas^2)$.
\end{quotation}
\end{titlepage}

\section{Introduction}
\renewcommand{\arraystretch}{2}                             %

Since experiments at the $e^+e^-$
 storage ring LEP started data taking a few
years ago, more than $7$  million events
have been collected at the $Z$ resonance
\cite{LEP}.
Besides the electroweak sector of the
Standard Model, LEP provides
an ideal laboratory for the investigation
of strong interactions. Due to their purely
leptonic initial state events are very
clean from both theoretical and
experimental point of view and represent
the ``golden'' place for testing
QCD. From cross section measurements
as well as the
analysis of event topologies the strong
coupling constant can be extracted.
Other observables which are measurable
with very high precision are the (partial)
$Z$ decay rates into hadrons
$\Gamma_{{\rm had}}/\Gamma_e=
20.763\pm 0.049$
and bottom quarks
$\Gamma_{b\bar{b}}/\Gamma_{{\rm had}}=
0.2200\pm 0.0027$ \cite{LEP}.
The program of experimentation at LEP  is
still not completed. The prospect of additional
$60$ $pb^{-1}$  per experiment
 means for example
that the relative uncertainty of the
partial decay rate into $b$ quarks
 $\Delta \Gamma_b/\Gamma_b$ may become even
smaller than one percent and an
experimental error for $\alpha_s$ of $0.002$
may be achieved \cite{Sch93}.

Also at lower energies significant
improvements can be expected in the
 accuracy of cross section measurements.
The energy region of around $10$ GeV
just below the $B\bar{B}$ threshold
will be covered with high statistics
at future $B$ meson factories. The cross
section between the charm and bottom
thresholds can be measured at the  BEPC
storage ring.
These measurements could
provide a precise value for $\alpha_s$
and --- even more important ---
a beautiful proof of the running of the
strong coupling constant.

In view of this experimental situation
 theoretical
predictions for the various observables
with comparable or even better
accuracy  become mandatory and
higher order radiative corrections
are required.
Significant improvements in these calculations
have been achieved in the last years and will
  be reviewed in this talk.
The discussion includes
singlet as well as nonsinglet corrections to the
vector and the axial vector correlator.
The corresponding ratios $R^V$ and $R^A$
enter both the formulae for the $e^+e^-$
annihilation cross section and the hadronic
$Z$ boson width.
Estimates of the theoretical uncertainties,
based on a variation of the renormalization
scale, will be reviewed in Section 2.
Particular emphasis is put on
 mass corrections from bottom
quark final states.
The quadratic mass corrections are
of relevance for
$Z$ decays and at lower energies. Quartic mass terms,
however, are of particular importance for the
low energy region, especially below and above
the $b\bar{b}$ threshold.
Corrections of order $m^2/s$ and new results
of order $\alpha_s^2 m^4$ \cite{CheKue94}
will be reviewed in Section 3.
The precise measurement of the rate
$\Gamma(Z\rightarrow b\bar{b})$ allows for a
determination of the mass of the top through its
impact on the effective $Zb\bar{b}$ vertex.
Secondary $b\bar{b}$ radiation as well as the
assignment of singlet contributions are covered
in Section 4.

Despite the remarkable confirmation of the
Standard Model by the precision tests at
LEP and SLC, experimental evidence for the
Higgs boson is still missing. Future colliders
like LHC and NLC may provide enough energy for
the production of the Higgs particle and
open the experimental door to the
electroweak symmetry breaking sector of the
Standard Model.
In Section 5 QCD corrections
to the Higgs decay rate  are discussed.
With $\Gamma(H\rightarrow b\bar{b})$
being the dominant decay channel
for intermediate Higgs masses,
second order singlet corrections to this
quantity  are calculated in
the heavy top mass limit. Leading quark mass
terms contribute to the same
order as nonsinglet graphs in the massless
limit and are  numerically sensitive
on the ratio $M_H^2/m_b^2$.

\section{The $Z$ Boson Width}
\subsection{Results for Massless Quarks}
Higher order QCD corrections
to $e^+e^-$ annihilation into hadrons
 were first
 calculated for the electromagnetic case in the
   the approximation of massless quarks.
Considering the annihilation process through
the $Z$ boson,
 numerous new features and subtleties
become relevant at the
 present level of precision.
\ice{
\begin {figure}
\begin{tabular}{ll}
&
\end{tabular}
\caption {
Renormalization scale dependence of the nonsinglet
(a) and the axial singlet (b)
massless QCD corrections.
$(\alpha_s(M_Z^2)=0.12)$
}
\end{figure}
}
An important distinction, namely  ``nonsinglet''
versus ``singlet'' diagrams,
originates from two classes of diagrams
with intrinsically different topology and
resulting charge structure.
The first class of diagrams consists of
nonsinglet
contributions with one fermion loop coupled
to  the  external currents. All these amplitudes
are proportional
to the square of the quark charge.
QCD corrections corresponding to these
 diagrams contribute
a correction factor which is independent
from the
current under consideration.
Singlet contributions
arise from a second class of
 diagrams where two currents
 are coupled to two
different fermion loops and hence
can be cut into two parts
by cutting gluon lines only.
They cannot be assigned to the
 contribution from one
individual quark species.
In the axial vector and the vector case the first
 contribution of this type
arises in order $\ordas^2)$
and $\ordas^3)$ respectively.
Each of them gives rise to a charge structure
different from the nonsinglet terms.

The nonsinglet terms have been calculated in
\cite{CheKatTka79} and
\cite{GorKatLar91}
 to second and third order in $\alpha_s$
respectively. As shown in Figure 1a the scale
dependence of the result (evaluated in the
$\msbar$-scheme and for a renormaliation scale
$\mu^2$ between $s/4$ and $4s$) is reduced
considerably through the inclusion of higher
 order terms.
The dependence of the result
due to the renormalization scale is an
indicator for the influence of yet uncalculated
higher order QCD corrections.
The remaining variation of $R_{NS}$ of about
$\pm 2\cdot 10^{-4}$
 for $\alpha_s(M_Z^2)=0.12$
translates into an uncertainty in
 $\alpha_s$ of $\pm 0.7\cdot 10^{-3}$.
This error estimate is in itself strongly
dependent on the actual value of $\alpha_s$.
For $\alpha_s(M_Z^2)=0.135$ for example, it is
amplified to $\pm 0.001$. These numbers are
quite comparable to the effect of an
$(\alpha_s/\pi)^4$ term with a coefficient around
100 as advocated in  \cite{KatSta94}.

 Similar observations apply for
 the singlet contributions to the axial current
correlator (see Figure 1b).
The order $\ordas^2)$ term
\cite{KniKue}
 still exhibits a
sizable scale dependence. Inclusion of the
  $\ordas^3)$ term
\cite{CheTar}, however,
leads to fairly stable answer.
It should be emphasized that the contribution of
this singlet term, which is effectively present
in the axial $b\bar{b}$ rate only, is larger
than the total contribution of the $\alpha_s^3$
term.

\section{Mass Corrections}
\subsection{Quadratic Mass Corrections}
The calculation of higher order QCD corrections
with massive quarks
 may be simplified
for small masses by expanding in
$m^2/s\ll 1$.
The operator product expansion of current
correlators, including subleading terms,
provides the theoretical framework.
It has been developed in
\cite{CheSpiGor85,CheSpi88} and applied to the
present problem in \cite{CheKue90,CheKueKwi92}.
 The expansion is
simultaneously performed in $\alpha_s$ and the
quark mass:
\beq
\ba{ll}\dsp
R^{V/A}
& \dsp
= R^{(0)} +
\frac{\bar{m}^2}{s}R^{(1)}_{V/A}
+\left( \frac{\bar{m}^2}{s}\right)^2 R^{(2)}_{V/A}
\\  & \dsp
=1+\api+\dots
+\frac{\bar{m}^2}{s}\left[
c_0^{V/A}+c_1^{V/A}\api+\dots \right]
+\left(\frac{\bar{m}^2}{s}\right)^2\left[
d_0^{V/A}+d_1^{V/A}\api+\dots \right]
\ea
\eeq
\ice{
\begin {figure}
\begin {tabular}{ll}
&
\end {tabular}
\caption {
Mass Corrections to $R^{(1)}_V$ (left) and $R^{(1)}_A$ (right).
The left bar represents the result in the on-shell-scheme, the
right one is obtained in the $\msbar$-scheme.
}
\end{figure}
}
The calculation is conveniently performed in the
$\msbar$-scheme \cite{MSbar} with the running mass, which has
the remarkable property that no logarithms of the
large ratio $s/m^2$ appear in $c_i$ for
arbitrary orders in $\alpha_s$. (This does not
hold true for the coefficients $d_i$.)
For the vector induced rate
the first coefficient $c_0^V$ vanishes and the
corrections $c_1^V,c_2^V,c_3^V$
have been calculated
in \cite{massless},
\cite{GorKatLar86}\footnote{Note that an
erratum for the $\zeta(3)$ term turned out to
be wrong and the originally published result
proved to be correct (see \protect\cite{Chetyrkin93}).} and
\cite{CheKue90} respectively.
For $n_f=5$ one obtains
\beq
R^{(1)}_V
=  12\frac{\bar{m}_b^2}{s}\api
\left\{
1+8.74\api+45.15\left(\api\right)^2
\right\}
\eeq
In Figure 2 it is demonstrated that the
calculation is fairly stable and higher order
corrections are small in the $\msbar$-scheme,
in contrast to the large changes in the
on-shell scheme.

Similar considerations apply for the axial
 current induced two point function. In
particular one can again demonstrate
 the absence of
large logarithms. The expansion starts in this
case with $c_0^A$ already. Nonsinglet terms have
been evaluated in \cite{CheKueKwi92}
  to order $\ordas^2)$,
the leading singlet terms of order $\ordas^2)$
were calculated in \cite{CheKwi93a,CheKwi93b}:
\beq\ba{ll}\dsp
R^{(1)}_A =
& \dsp
  -6\frac{\bar{m}_b^2}{s}\api
\left\{
1+\frac{11}{3}\api
+\left(\api\right)^2
\left(11.296+\ln\frac{s}{m_t^2}\right)
\right\}
\\ & \dsp
  -10\frac{\bar{m}_b^2}{m_t^2}\left(\api\right)^2
\left(\frac{8}{81}-\frac{1}{54}\ln\frac{s}{m_t^2}\right)
\ea\eeq
As shown in Figure 2
the calculation in the on-shell scheme
exhibits significant changes with the inclusion
of coefficients with large logarithms.  In the
$\msbar$-scheme the expansion is remarkably
stable. The size of the corrections is comparable
to the anticipated experimental precision.

\subsection{Quartic Mass Corrections}
The second order  calculation of quartic
 mass corrections
 presented below is based on
\cite{CheKue94}.
The operator product expansion included power
law suppressed terms up to operators of
dimension four induced by
 nonvanishing quark masses.
Renormalisation group arguments
 similar to those employed already in
\cite{CheKue90,CheKueKwi92}
allowed to deduce the $\as^2 m^4$ terms. The
calculation was
 performed for vector and axial vector
current nonsinglet
correlators. The first one is of course
relevant for electron positron annihilation
 into heavy quarks at
arbitrary energies, the second one for $Z$
decays into $b$ quarks and
for top production at a future linear collider.
Below only the results for $R_V$ will be presented.

QCD corrections to the vector current correlator
in order $\as$ and for arbitrary $m^2/s$ were
derived in \cite{massless}. These
are conventionally expressed in
terms of the pole mass
denoted by $m$ in the following. It is
straightforward to expand these
results in $m^2/s$ and one obtains
\begin{eqnarray}
R_V&=&1 - 6\frac{m^4}{s^2} - 8\frac{m^6}{s^3}\\
&& +\frac{\as}{\pi}\biggl[1+12 \frac{m^2}{s}
+ \left( 10 - 24\ln (\frac{m^2}{s})\right)\frac{m^4}{s^2}
                    \nonumber\\
   &&  -\frac{16}{27}\left(47 + 87 \ln (\frac{m^2}{s})\right)
   \frac{m^6}{s^3}\biggr] \nonumber
{}.
\label{1}
\end{eqnarray}

The approximations to
the correction function for the vector
 current correlator
(including successively higher orders
and without the
factor $\alpha_s/\pi$) are compared to
the full result in
Fig.\ref{kvexp}.
As can be seen in Fig.\ref{kvexp},
for high energies, say for $2 m_b/\sqrt{s}$
 below 0.3, an
excellent approximation is provided by the
constant plus the $m^2$ term.
In the region of $2m/\sqrt{s}$ above 0.3
the $m^4$ term becomes
increasingly important. The inclusion of
 this term improves the
agreement significantly and leads to an
excellent approximation even up
to $2m/\sqrt{s}\approx 0.7$ or 0.8. For the
narrow region between 0.6
and 0.8 the agreement is further improved
 through the $m^6$ term.
The logarithms accompanying the $m^4$ terms
can also be absorbed through a redefinition of $m$ in terms of
the $\overline{MS}$ mass \cite{Tarrach81,Narison87}
at scale $s$
and one obtains
($\ovl{m} \equiv \ovl{m}(s)$)
\ice{
\begin{figure}
\begin{center}
\leavevmode
\caption{\label{kvexp}Comparison between the
complete ${\cal O}(\alpha_s)$ correction
function (solid line) and approximations
 of increasing order
(dashed lines) in $m^2$ for vector
induced rates.}
\end{center}
\end{figure}
}
\begin{eqnarray}
R_V& =& 1 - 6\frac{\ovl{m}^4}{s^2}
 -8\frac{\ovl{m}^6}{s^3}\\
&&+\frac{\as}{\pi}\left
    [1 + 12\frac{\ovl{m}^2}{s}
- 22 \frac{\ovl{m}^4}{s^2}
-  \frac{16}{27}\left(6\ln(
 \frac{\ovl{m}^2}{s})  + 155\right)
  \frac{\ovl{m}^6}{s^3}
   \right] \nonumber
. \nonumber
\end{eqnarray}
This resummation is possible
 for the second and fourth powers of $m$
in first
order $\as$ and in fact for $m^2$
corrections to $R_V$ and $R_A$ in all
orders of $\as$. However, logarithmic
terms persist in the
$m^4$ corrections, starting from ${\cal O}(\as^2)$.

Motivated by the fact that the first few
terms in the $m^2/s$ expansion
provide already an excellent
 approximation to the complete answer in
order $\as$, the three loop
corrections have been calculated.
The calculation is based on
 the operator product expansion of
the T-product of two vector currents
$J_\mu=\ovl u\g_\mu d $.
Here $u$ and $d$ are just two generically
 different quarks with
masses $m_u$ and $m_d$. Quarks which are
 not coupled to the external
current will influence the result in order
 $\as^2$ through their
coupling to the gluon field.
The result may be immediately transformed
to the case of the
electromagnetic current of a heavy, say,
$t$ (or $b$ ) quark.

\ice{
\begin{figure}
\leavevmode
\caption{\label{alphasexp}Contributions
to $R^V$ from $m^4$ terms
including successively higher
orders in $\alpha_s$ (order $\alpha_s^0$/
 $\alpha_s^1$/ $\alpha_s^2$
corresponding to dotted/ dashed/ solid lines)
as functions of
$2m_{\rm pole}/\protect\sqrt{s}$.}
\end{figure}
The asymptotic behaviour of the transverse part
of this
(operator valued) function for
  $Q^2 = -q^2\to\infty$  is given by
an OPE  of the following form.
(Different powers of $Q^2$ may be studied
separately and only operators of dimension
4 are displayed.)
\beq
i\int T(J_\mu(x) J^+_\nu(0))\ex^{iqx} \dif x
\bbuildrel{=}_{q^2\to\infty}^{} \frac{1}{Q^4}
\sum_n
(q_\mu q_\nu -g_{\mu\nu} q^2)
\fos{T}{C_n}(Q^2,\mu^2,\alpha_s)
O_n+\dots
\EQN{142}
\eeq
Only the gauge invariant operators
$G_{\mu\nu}^2, m_i\bar{q}_jq_j$ and a polynomial
of fourth order in the masses contributes to
physical matrix elements. The coefficient
functions were calculated in \cite{CheSpiGor85},
the vacuum expectation values of the relevant
operators in \cite{Bro81,CheSpi88,BraNarPic92}.
Employing renormalization group arguments
the vacuum expectation value of $\sum_n C_nO_n$
is under control up to terms of order $\alpha_s$
as far as the constant terms are concerned
and even up to $\alpha_s^2$ for the
logarithmic terms proportional to $\ln Q^2/\mu^2$.
Only these logarithmic terms  contribute to
the absorptive part. Hence one arrives at the
full answer for $\alpha_s^2m^4/s^2$ corrections.
Internal quark loops contribute in this order,
giving rise to the terms proportional
to $\sum m_i^2$ and $\sum m_i^4$ below.
\begin{figure}
\leavevmode
\caption{\label{massexp}Predictions for
 $R^V$ including successively higher
orders in $m^2$.}
\end{figure}
}

 The result reads (below we set for brevity
the $\ovl{\mbox{\rm MS}}$ normalization
scale $\mu = \sqrt{s}$
and $\ovl{m}_u(s) = \ovl{m}_d(s) = \ovl{m}$;
$a\equiv \as/\pi$)
\begin{eqnarray}
\dsp
R_V & = & 1 + O(\ovl{m}^2/s)
 -6 \frac{\ovl{m}^4}{s^2}\left(1 + \frac{11}{3} a \right)
\nonumber
\\
&&
\dsp
+
a^2\frac{\ovl{m}^4 }{s^2} \left[
f \left(
\frac{1}{3}  \ln\left(\frac{\ovl{m}^2}{s}\right)
-1.841
  \right)
\right.
\nonumber
\\
&&
\left.
\dsp
-\frac{11}{2} \ln\left(\frac{\ovl{m}^2}{s}\right)
+ 136.693
+12  \sum_i \frac{\ovl{m}_i^2}{\ovl{m}^2}
\right.
\nonumber
\\
&&
\left.
\dsp
 - 0.475   \sum_i \frac{\ovl{m}_i^4}{\ovl{m}^4}
 -
\sum_i \frac{\ovl{m}_i^4}{\ovl{m}^4}
\ln\left(\frac{\ovl{m_i}^2}{s}\right)
\right],
\EQN{VM4N}
\end{eqnarray}

Note that the sum over $i$ includes also the quark
coupled to the external current and with
 mass denoted by $m$.
Hence in the case with one heavy quark $u$  of
mass $m$ ($d \equiv u)$ one should set
$\sum_i \frac{\ovl{m}_i^4}{\ovl{m}^4} = 1$
and
$\sum_i \frac{\ovl{m}_i^2}{\ovl{m}^2} = 1$.
In the opposite case when one considers
the correlator of light
(massless) quarks the heavy quark appears only
through its coupling to gluons. There one finds:
\beq
\dsp
R_V  = 1 +   O(\ovl{m}^2/s)
+
a^2 \frac{\ovl{m}^4 }{s^2} \left[
\frac{13}{3}
-\ln\left(\frac{\ovl{m}^2}{s}\right)
-4 \zeta(3)
\right].
\EQN{VAM4}
\eeq

The $Z$ decay rate is hardly affected by
the $m^4$ contributions. The
lowest order term evaluated with $\ovl{m}=2.6
 \ {\rm GeV}$ leads to a
relative
suppression (enhancement) of about
 $5\times 10^{-6}$ for the vector
(axial vector) current induced
 $Z\to b\bar b $ rate.
Terms of increasing order in $\alpha_s$
 become successively smaller. It
is worth noting, however, that the
 corresponding series, evaluated in
the onshell scheme, leads to terms which
are larger by about one order
of magnitude and of oscillatory signs.
 The $m_b^4$ correction
to $\Gamma(Z\to q\bar q)$ which starts in
 order $\alpha_s^2$
is evidently even smaller.
{}From these
considerations it is clear that $m^4$
corrections to the $Z$ decay
rate are well under control --- despite
 the still missing singlet piece
--- and that they can be neglected for
 all practical purposes.

The situation is different in the low
energy region, say
several GeV above the charm or the bottom
threshold. For  definiteness
the second case will be considered and
for simplicity
all other masses will be put to
zero
The contributions to $R^V$ from $m^4$ terms are
presented  in Fig.\ref{alphasexp}
as functions of $2m/\sqrt{s}$ in the range
 from 0.05 to 1.
As input parameters
$m_{\rm pole}=4.7{\rm GeV}$ and
$\alpha_s(m_Z^2)=0.12$ have
 been chosen.
Corrections of higher orders are added
 successively. The prediction is
fairly stable with increasing order in
 $\alpha_s$ as a consequence of
the fact that most large logarithms were
absorbed in the running mass.
The relative magnitude of the sequence
 of terms from the $m^2$
expansion
is displayed in Fig.\ref{massexp}. The
curves for $m^0$ and $m^2$  are based on
corrections up to third order in $\alpha_s$
 with the $m^2$ term starting
at first order. The $m^4$ curve receives
 corrections from order zero to
two.

Of course, very close to threshold, say
above 0.75 (corresponding to
$\sqrt{s}$ below 13 GeV) the approximation
is expected to break down,
as indicated already in Fig.\ref{kvexp}.
Below the $b \bar b$ threshold, however, one
may decouple the bottom quark and consider
 mass corrections from the
charmed quark within the same formalism.

Also $R_{q\bar q}$ where $q$ denotes a
 massless quark  coupled to the
external current is affected by virtual or
 real heavy quark radiation.
The $m^2$ terms have been calculated in
  \cite{CheKue90} and start
from order $\alpha_s^3$:
\beq
\delta R=-\left(\frac{\alpha_s}{\pi}\right)^3
\frac{4\ovl{m}^2}{s} (15-\frac{2}{3} f)
(\frac{4}{3} -\zeta(3) )
\eeq
The terms of order $\alpha_s^2m^4$ were
 given above. Both lead to
corrections of ${\cal O}( 10^{-4})$,
 (evaluated at an
energy $\sqrt{s}$ of 10 GeV) and can
 be neglected for all practical
purposes.

\section{Secondary $b\bar{b}$ Production}
The formulae for the QCD corrections to the total
 rate $\Gamma_{{\rm had}}$
 have a simple, unambiguous meaning. The
theoretical predictions for individual $q\bar{q}$
channels, however, require additional
interpretation. In fact, starting from order
${\cal O}(\alpha_s^2)$ it is no longer possible
to assign all hadronic final states to well
specified $q\bar{q}$ channels in a unique manner.
For definiteness we shall try to isolate and define
$\Gamma(Z\rightarrow b\bar{b})$.
The thorough understanding of QCD and mass
corrections is mandatory to disentangle weak
corrections with a variation of about 1\% for
$m_t$ between 150 and 200 GeV from QCD effects.

\vspace{2ex}
{\bf 1. {\it Non Singlet Contribution}}

\noindent
The vector and axial vector induced rates receive
(non singlet) contributions from the diagrams,
 where the heavy quark pair is radiated off a
light $q\bar{q}$ system.
 The rate for this
 specific contribution to the $q\bar{q}b\bar{b}$
final state is given by \cite{HoaJezKueTeu94}
\beq
\ba{ll}\dsp
R^{NS}_{q\bar{q}b\bar{b}}
=
& \dsp
\frac{\Gamma^{NS}
_{q\bar{q}b\bar{b}}}
{\Gamma^{{\rm Born}}_{q\bar{q}}}
=
\left(\frac{\alpha_s}{\pi}
 \right)^2
\frac{1}{27}\left[
\ln^3 \frac{s}{m_b^2} - \frac{19}{2} \ln^2 \frac{s}{m_b^2}
 + \left(\frac{146}{3}-12\zeta(2)
   \right) \ln \frac{s}{m_b^2}
 \right.
\\ &\dsp
\left.
-\frac{2123}{18}+38 \zeta(2) + 30\zeta(3)
\right]
\\ & \dsp
=\left(\frac{\alpha_s}{\pi}
 \right)^2\cdot
(0.922/0.987/1.059)
\;{\rm for}\; m_b=(4.9/4.7/4.5)\;{\rm GeV}
\ea
\eeq

Despite the fact that $b$ quarks are present
in the four fermion final state,
the natural prescription is to assign
these contributions to the rate into the
$q\bar{q}$ channel. Therefore those events
with primary light quarks and secondary bottom
quarks must be subtracted experimentally
from the partial rate $\Gamma_{b\bar{b}}$. This
should be possible, since their signature is
characterized by a large invariant mass of
the light quark pair and a small invariant mass
of the bottom system, which is emitted
collinear to the light quark momentum.
If this subtraction is not performed, the
$b\bar{b}$ rate
is overestimated by
 (for $m_b=4.7$ GeV and $\alpha_s=0.115 \dots 0.18$)
\beq
\Delta \equiv
\frac{\dsp \sum_{q=u,d,s,c}\Gamma^{NS}_{q\bar{q}b\bar{b}}}
{\Gamma_{b\bar{b}}}
\approx 0.005\dots 0.013
\eeq

\vspace{2ex}
{\bf 2. {\it Singlet Contribution}}

\noindent
The situation is more complicated for the four fermion
final state from the singlet contribution
which originates from the interference
term $R^S_{q\bar{q}b\bar{b}}$
between the $q\bar{q}$ and $b\bar{b}$ induced
amplitudes.
It cannot be assigned in an unambiguous
way to an individual $q\bar{q}$ partial
rate.

For the vector current induced rate
and after phase space integration this term
vanishes as a consequence of charge conjugation
(Furry's theorem).

For the axial current induced rate, however,
this interference term gives a nonvanishing
contribution which remains finite even in the
limit $m_q \rightarrow 0$. Neglecting
all masses one obtains
$$
R^{S}_{b\bar{b}b\bar{b}}
= - \frac{1}{2}R^{S}_{b\bar{b}u\bar{u}}
= + \frac{1}{2}R^{S}_{b\bar{b}d\bar{d}}
= - \frac{1}{2}R^{S}_{b\bar{b}c\bar{c}}
= + \frac{1}{2}R^{S}_{b\bar{b}s\bar{s}}
= -0.153 \left(\frac{\alpha_s}{\pi}\right)^2
$$
Among the final states from these
singlet diagrams with at least one $b\bar{b}$ pair,
only the $b\bar{b}b\bar{b}$ term remains after all
compensations have been taken into account.
Numerically it is tiny. Let us compare it with the
other singlet contributions consisting of
$b\bar{b}(g)$ jets. They
 can be assigned to
$\Gamma_{b\bar{b}}$ in a unique way, although they
are induced by the $b\bar{b}$ and the $t\bar{t}$
currents. These two cuts, in particular the
leading term with two $b\bar{b}$ jets, dominate
the four fermion final  state by
 a factor of $20$.
Therefore it is suggestive to assign the total
singlet contribution to $\Gamma_{b\bar{b}}$.

\section{The Higgs Decay Rate}
The  partial width
$\Gamma(H\rightarrow b\bar{b})$ is
significantly affected by QCD
radiative  corrections.
First order $\ordas)$ corrections
 were studied in
\cite{higgs}.
Besides the overall
$m_b^2$ dependence
 due to the Higgs-bottom coupling,
the otherwise massless corrections in second
 order
${\cal O}(\alpha_s^2)$ were obtained
by \cite{GorKatLarSur8490}. The resulting expression
for the decay rate in question reads
\beq
\EQN{bb1}
\Gamma(H\rightarrow b\bar{b})
=\frac{3 G_F}{4\sqrt{2}\pi} M_H \bar{m}_b^2
\left[
1 + \Delta\Gamma_1 \left(\api\right) +
 \Delta\Gamma_2 \left(\api\right)^2
\right],
\eeq
with
$\Delta\Gamma_1   = \frac{17}{3}, \ \
  \Delta\Gamma_2  = 29.14
$.

In this talk we present the calculation of
an additional contribution
to the ``massless'' ${\cal O}(\alpha_s^2)$
 corrections, which were neglected
in the literature so far.
These
 singlet corrections
are due to the ``light-by-light'' type diagram
 which we  calculate
for nonvanishing quark masses
 in the heavy top limit.
The Yukawa couplings of the fermions
yield  a factor $m_bm_t$ and
 each fermion trace
  produces a factor $m_b$ and
$m_t$ respectively. Together with
a power supression of $1/m_t^2$ on
dimensional grounds all mass factors
 are combined to $m_b^2$, which
is of the same
order as the otherwise
massless $\ordas^2)$ nonsinglet corrections.

The absorptive part of the triangle diagram
is given by
\beq
\Delta\Gamma^{abs}
= \frac{3 G_F}{4\sqrt{2}\pi} M_H m_b^2
\left(\api\right)^2\frac{1}{3}
\left[\frac{28}{3}-2\ln
 \frac{M_H^2}{m_t^2}\right]
{}.
\eeq
Here also the decay mode into two gluons is
included. It is separately known
 \cite{gg} and should  be subtracted
in order to arrive at the singlet correction for
the bottom final state:
\beq
\Delta\Gamma^S_{H\rightarrow b\bar{b}}
= \frac{3 G_F}{4\sqrt{2}\pi} m_b^2 N_C
\left(\api\right)^2\frac{1}{3}
\left[8-\frac{\pi^2}{3}-2\ln \frac{M_H^2}{m_t^2}
+\frac{1}{3}\ln^2\frac{m_b^2}{M_H^2}\right]
{}.
\EQN{bb3}
\eeq
At last, combining (\ref{bb1}) and (\ref{bb3}) we get
\beq
\ba{c}
\dsp
\Gamma(H\rightarrow b\bar{b}) =
\\
\dsp
\frac{3 G_F}{4\sqrt{2}\pi} M_H \bar{m}_b^2
\left[
1 + \frac{17}{3} \left(\api\right) +
 (
      30.71 - \frac{2}{3}\ln\frac{M_H^2}{m_t^2}
   +  \frac{1}{9}\ln^2\frac{\bar{m}_b^2}{m_H^2}
 )
 \left(\api\right)^2
\right].
\ea
\EQN{bb4}
\eeq
It should be finally stressed that
the separation between the decay channels into
bottom quarks and gluons leads to a
logarithmically enhanced $\ln^2 (m_b^2/M_H^2)$
correction to the decay width.

\end{document}